# Spectral intensity and phase changes of short laser pulses by two-photon absorption

Tsogvoo Khos-Ochir[1], Kim Myung-Whun[1,2] and Purevdorj Munkhbaatar[1,3,*]

[1]Institute of Photonics and Information Technology, Chonbuk National University, Jeon-ju 54896, Republic of Korea
[2]Department of Physics, Chonbuk National University, Jeon-ju 54896, Republic of Korea
[3]Department of Physics, National University of Mongolia, Ulaanbaatar 210646, Mongolia
*corresponding author: p_munkhbaatar@num.edu.mn

**ABSTRACT**

We calculated third order non-linear polarization to estimate the two-photon absorption of non-interacting two-level molecules in the transmission-type degenerate pump-probe geometry. The spectral intensity and the phase changes of the laser pulses when passing through a thin dielectric slab composed of the molecules were considered. We also investigated the effect of the decay rate of the molecules and the chirp of the pulses on their spectral intensity and phase changes.

## 1. INTRODUCTION

Two-photon absorption (TPA) has attracted great attention because of its application to microscopic imaging [1]. Two-photon imaging was developed to detect fluorescence signals at the geometrical focus of a microscopy system [2,3]. The studies involving TPA have mostly involved investigations of the intensity dependence of the signal in sample materials [4,5]. However, as interest in deep tissue imaging in optically thick biological samples has increased in recent years, information on the refractive index changes during the TPA process is important because it is a key factor in image formation and aberration correction [6,7]. The absorption coefficient and the refractive index are connected to the real and the imaginary part of the complex susceptibility function of a material. Usually, the detection of the information associated with the real part is difficult, but experimental techniques for obtaining phase information of the short laser pulses have been developed [8,9]. These techniques can stimulate studies related to the refractive index or phase change information during the TPA process.

Few theoretical studies have dealt with the spectral and time-resolved complex susceptibility for TPA. This is possibly due to the fact that TPA is a third order nonlinear process in which four light fields are intertwined in complicated ways with various time intervals. Some proposed mathematical models in the literature primarily deal with the case of monochromatic excitation [10-12], while current studies of TPA use ultra-short laser pulse possessing with very broad spectral bandwidth. It is also necessary to consider the effect of higher order spectral phase for short pulses. Fortunately, the TPA process is usually used to excite small fluorescent molecules. As such, the realistic modeling of the electronic structure of the molecules may not be a critical aspect. A theoretical guideline regarding a simple ideal molecule is worthy to compare with the intensity and phase change of ultra-short laser pulses with higher order phase as measured experimentally.

In this paper, we provided a theoretical guideline to examine the properties of TPA signals in regard to intensity and phase spectra of ultra-short laser pulses. We estimated the spectrally and temporally resolved TPA signals for a thin dielectric slab sample composed of two-level non-interacting molecules in the degenerate pump-probe geometry, which is widely used in many experiments. Further, in many nonlinear optical experiments using short laser pulses, the second-order phase of the pulses, i.e., chirp, often varies as they pass through thick transparent samples. We calculated the effect of pulse chirp on the spectrally and temporally resolved intensity and phase function of short laser pulses.

## 2. THEORY AND MODEL

We consider a thin slab of a dielectric material as the sample. This material is composed of non-interacting two-level atoms. The energy difference between the ground state ($|g\rangle$) and the final state ($|f\rangle$) of the atom is close to the twice the light frequency. We consider a virtual intermediate state (($|e\rangle$)). The energy difference between the ground state and the intermediate state is similar to the light frequency. The one-photon transition between $|g\rangle$ and $|e\rangle$ is not allowed, but the two-photon transition occurs via the $|e\rangle$ state. The material is transparent at the light source frequency but opaque at twice the frequency.

Firstly, we obtain the third order polarization $P^{(3)}$ of the sample material. We treat the laser light as a classical electric field $E(t)$. The coupling Hamiltonian between the light and matter is $H_{int} = -\mu E(t)$, where $\mu$ is the dipole operator. The $P^{(3)}$ is determined as following [13]:

$$P^{(3)}(t) = \sum_{m=0}^{3} \langle \psi^{(3-m)}(t)|\mu|\psi^m(t)\rangle. \quad (1)$$

Here $\langle \psi^{(3-m)}(t)|$ and $|\psi^m(t)\rangle$ are the perturbed bra and ket vectors of the material system to the $(3-m)$-th and $m$-th order in the external field.

To count the time ordering of the electric fields in Eq. (1), we used Keldysh-Schwinger loop diagrams as shown in Figs. 1(a)-1(d) [14]. The system interacts with the laser fields $E_{pu}(t)$, $E_{pr}(t)$ and $E_{pu}^*(t)$ at times $\tau_1$, $\tau_2$ and $\tau_3$, respectively. $E_{pu}(t)$ is the pump-field and $E_{pr}(t)$ is the probe-field. The polarization is calculated at $\tau_4$ by integrating over the time variables $\tau_j$ ($j = 1,2,3,4$). Figure 1 represents all possible time orderings of $\tau_j$ along the loop. Eq. (1) can be simply represented by introducing the Green functions. The third order polarization corresponding to the TPA can be written as follows:

$$P^{(3)}(\omega_s, \tau) = \mu_{ge}\mu_{ef}\mu_{fe}\mu_{eg} \iint d\omega_1 d\omega_2 [E_{pu}(\omega_1)E_{pr}(\omega_2)E_{pu}^*(\omega_1+\omega_2-\omega_s)e^{-i(\omega-\omega_2)\tau}G_e(\omega_1)G_f(\omega_1+\omega_2)G_e(\omega_s)$$
$$+E_{pu}(\omega_2)E_{pr}(\omega_1)E_{pu}^*(\omega_1+\omega_2-\omega_s)e^{-i(\omega-\omega_1)\tau}G_e(\omega_1)G_f(\omega_1+\omega_2)G_e(\omega_s)$$
$$-E_{pu}(\omega_1)E_{pr}(\omega_2)E_{pu}^*(\omega_1+\omega_2-\omega_s)e^{-i(\omega-\omega_2)\tau}G_e(\omega_1)G_f(\omega_1+\omega_2)G_e^+(\omega_1+\omega_2-\omega_s)$$
$$-E_{pu}(\omega_2)E_{pr}(\omega_1)E_{pu}^*(\omega_1+\omega_2-\omega_s)e^{-i(\omega-\omega_1)\tau}G_e(\omega_1)G_f(\omega_1+\omega_2)G_e^+(\omega_1+\omega_2-\omega_s)].$$
(2)

Here, $\omega_s$ is the frequency of the TPA signal. Note that $\tau$ represents the time delay between $E_{pu}(t)$ and $E_{pr}(t)$, which should be distinguished from the interaction time $\tau_j$. $G_\eta(\omega) = 1/(\omega - \omega_{\eta g} + i\Gamma_{\eta g})$ is the retarded Green's function and $G_\eta^+(\omega) = 1/(\omega - \omega_{\eta g} - i\Gamma_{\eta g})$ is the advanced Green's function. $\mu_{\eta g}$ is the dipole matrix element, $\omega_{\eta g}$ is the frequency difference, and $\Gamma_{\eta g}$ is the dephasing constant between $|g\rangle$ and $|\eta\rangle$ ($\eta = e, f$). These parameters can be considered as constants in the fitting. The details of the calculation for Eqs. (1) and (2) can be found in the literature [14].

Next, we estimate the change of the light field when the field passes through the thin dielectric slab in a conventional degenerate pump-probe geometry. We assume that the light propagates along the z-axis in the positive direction. The slab is at $0 < z < l$, and $k_{pr}l \gg 1$. Here, $l$ is the thickness of the slab and $k_{pr}$ is the wavenumber of the probe light field. The interaction between the light and the material induces a nonlinear polarization $P_{NL}(z)$ inside the material, which serves as the light source of the signal field. The relation between the probe light field ($E_{pr}(z, \omega_s)$) and $P_{NL}(z, \omega_s)$ can be represented using Maxwell's equation for a slowly varying amplitude approximation as follows [15]:

$$\frac{\partial E_{pr}(z,\omega_s)}{\partial z} = i\frac{2\pi\omega_s}{c \cdot n(\omega_s)} P_{NL}(z,\omega_s). \quad (3)$$

Here $n(\omega_s)$ and $c$ are the refractive index of the material which acts on the probe field and the speed of light in a vacuum, respectively. The probe field can be represented by its amplitude and phase: $E_{pr}(z,\omega_s) = E_0(z,\omega_s)e^{-i\phi(z,\omega_s)}$. Substituting for the probe field in Eq. (3) and multiplying by $E_{pr}^*(z,\omega_s)$ results in the coupled equation for the intensity $I$ and phase $\phi$:

$$\frac{\partial I}{\partial z} = -I \cdot \frac{4\pi\omega_s}{c \cdot n(\omega_s)} Im[P_{NL}(z,\omega_s)/E_{pr}^*(\omega_s)], \quad (4)$$

$$\frac{\partial \phi}{\partial z} = \frac{2\pi\omega_s}{c \cdot n(\omega_s)} Re[P_{NL}(z,\omega_s)/E_{pr}^*(\omega_s)]. \quad (5)$$

Since the slab is thin, we can approximate $\partial I/\partial z \approx \Delta I/\Delta z$ and $\partial \phi/\partial z \approx \Delta \phi/\Delta z$. Here, $\Delta I$ ($\Delta \phi$) corresponds to change of the signal intensity (phase) before the pump pulse. $\Delta z$ can be considered as the thin slab thickness $l$. A conventional pump-probe experiments uses the self-heterodyne detection method, which measures the difference between the probe field and the reference field (or local oscillator). We are interested in obtaining the difference or change of the signals (i.e. $\Delta I$ and $\Delta \phi$). Therefore, we ignore the z-dependence of the polarization and collect the third order terms:

$$\Delta I = -\frac{4\pi\omega_s l}{c \cdot n(\omega_s)} Im[P^{(3)}(\omega_s) \cdot E_{pr}(\omega_s)], \quad (6)$$

$$\Delta \phi = \frac{2\pi\omega_s l}{c \cdot n(\omega_s)} Re[P^{(3)}(\omega_s)/E_{pr}^*(\omega_s)]. \quad (7)$$

By considering pulse $\sigma$-band widths much smaller than the pulse $\omega_0$-center frequency, the intensity change $\Delta I$ and phase change ($\Delta \phi$) are proportional to:

$$\Delta I(\omega_s, \tau) \propto -Im[P^{(3)}(\omega_s, \tau) \cdot E_{pr}(\omega_s)], \quad (8)$$

$$\Delta \phi(\omega_s, \tau) \propto Re[P^{(3)}(\omega_s, \tau)/E_{pr}^*(\omega_s)]. \quad (9)$$

Here, the third order polarization $P^{(3)}(\omega_s, \tau)$ is depending on $\omega_s$, the frequency of transmitted light through the material and $\tau$, the time delay between pump and probe pulses. We can estimate $\Delta I(\omega_s, \tau)$ and $\Delta \phi(\omega_s, \tau)$ by using the Eqs. (2), (8), and (9) if the laser pulse field is known.

## 3. RESULTS AND DISCUSSION

We simulated the TPA signal of a thin dielectric slab in the conventional laser pulse pump-probe experiment. For the simulation, we used a Gaussian-shape electric field for both the pump and probe fields of which the spectrum has the following functional form:

$$E(\omega) = E_0 e^{-\frac{(\omega-\omega_0)^2}{\sigma^2}} e^{-i\varphi(\omega)}. \quad (10)$$

Here $E_0$ is the amplitude of the Gaussian pulse field. $\sigma$ is the pulse width and $\omega_0$ is the center frequency of the pulse. $E_0 = 1$ and $\sigma = 0.25$ PHz for all the following results.

We calculated the intensity change ($\Delta I$) and phase change ($\Delta \phi$) in the case of $\omega_{fg} = 2\omega_0$, $\omega_0 = 2.35$ PHz, $\Gamma = 100$ PHz, and $\gamma = 10^{-6}$ PHz. $\Gamma$ is the dephasing rate ($\Gamma_{eg}$) and $\gamma$ denotes the population rate ($\Gamma_{fg}$). The results are shown in Figs. 2(a) and 2(b). The $\Delta I$ function shown in Fig. 2(a) has a perfectly symmetric shape in both temporal and spectral domains with respect to $\tau = 0$ and $\omega_s = 2.35$ PHz. The amplitude of the complex $\chi^{(3)}$ function determines the shape of the $\Delta I$ function, and the dipole moment $\mu$ determines the shape of $\chi^{(3)}$. However $\mu$ has no particular functional dependence in our simulation. The shape of the $\Delta I$ function should reflect the effect of the Green's functions, which is the simplest response of the material system to an external electric field. If the spectral features more are complicated than that represented in Fig. 2(a) in an experiment, the features are probably due to the complexity of $\mu$. The $\Delta \phi$ function shown in Fig. 2(b) looks like an odd function in both the temporal and spectral domains. Note that the spectral structure of the $\Delta \phi$ function is more complicated and the curvature of the function has a much steeper gradient. This means that $\Delta \phi$ can be more sensitive to small change than $\Delta I$ under usual circumstances.

We examined the effect of the change in the resonance frequency and the decay parameters. The change of the resonance frequency resulted in only a constant shift of the response function, but the change of the decay parameter presented some intriguing results. First, we examined the effect of the change of $\Gamma$. For $\gamma = 10^{-6}$ PHz and $\omega_0 = 2.35$ PHz, we evaluated three cases: $\Gamma = 100$, 50, and 5 PHz. The structure of the $\Delta I$ did not changed except for the magnitude. The ratio of the magnitude change was 1:0.25:0.0025, which means that $\Delta I$ is proportional to $\Gamma^{-2}$. The $\Delta \phi$ also exhibited changes only in their magnitudes. The ratio of the magnitude change was 1:0.4:0.004. This means that the stable (long lifetime) intermediate state enhanced the TPA process as expected. Next we examined the effect of $\gamma$. The $\Delta I$ did not show any significant change except for a change in magnitude. However, $\Delta \phi$ resulted in a non-trivial change in the higher order response. Figure 3 shows $\Delta\phi$ at $\omega_s = 2.5$ PHz as a function of time delays with $\gamma$, $10\gamma$, $100\gamma$, $1000\gamma$, and $10000\gamma$. Here, $\gamma = 10^{-6}$ PHz and the other parameters were the same as for Fig. 3(a). There are no changes in $\Delta\phi$ for $\gamma < 10^{-6}$ PHz. On the other hand, the intensity of $\Delta\phi$ decreased as $\gamma$ increased more than $10^{-6}$ PHz. If multiplied by some constants, $\Delta\phi(\gamma)$ and $\Delta\phi(10\gamma)$ are the same. However, for $\Delta\phi$ with $100\gamma$ or larger values, the curvature of the valleys and hills cannot be recovered by a simple multiplication. This means that a faster excitation decay causes smaller change in the third order phase.

In real experiments, the phase of the laser pulse can be chirped due to optical elements such as lenses and windows. The phase is usually positively chirped by transparent materials. We examined the possible effects of positive chirp on the $\Delta I$ and $\Delta\phi$ signals as shown in Figs. 4(a) and 4(b). The chirp was 10 fs$^2$, i.e. $\varphi(\omega) = 10(\omega - \omega_0)^2$. This caused the hill in the $\Delta I$ to be rotated. The chirp also caused the hills and valleys in $\Delta\phi$ to be squeezed and twisted. If the rotated or twisted structure is observed using frequency-time pump-probe spectroscopy, the chirp state of the incident laser pulse should be investigated.

## 4. CONCLUSION

We determined the intensity and phase change functions of a laser pulse for two-photon absorption. As an example, we assumed that the laser pulse was described by a Gaussian spectral function and was transmitted through a thin dielectric slab composed of two-level non-interacting molecules. The calculation results and the equations can be used as a theoretical guideline to analyze the intensity and the phase change information of TPA signals obtained from degenerate pump-probe experiments using broad-band ultrashort laser pulses.

**ACKNOWLEDGMENTS**

This research was supported by the National Research Foundation of Korea (Grants No. 2016R1D1A1B03934648 and No. 2016K2A9A1A09914398); by the National Research Foundation of Mongolia (Grant No. SSA 017/2016).

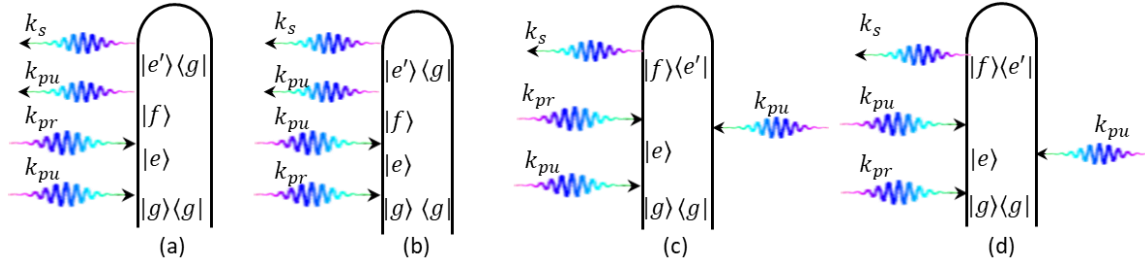

Fig. 1 Keldysh-Schwinger loop diagrams for the pathways of the two photon absorption in the pump-probe experiment. The diagrams are distinguished by the time ordered interactions of the pump ($k_{pu}$) and probe ($k_{pr}$) pulses with the material.

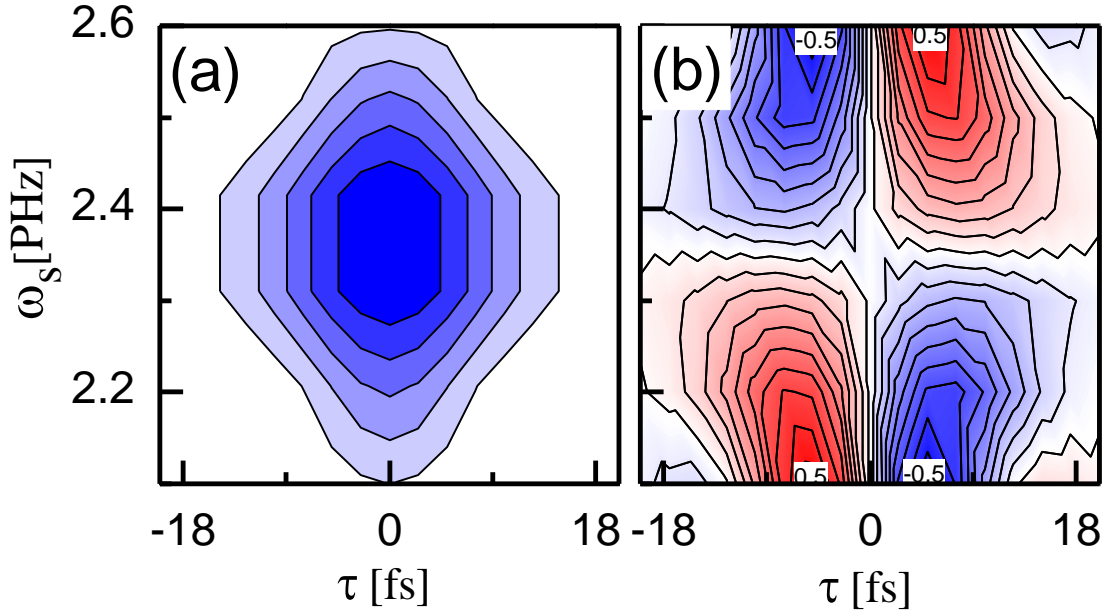

Fig. 2 Simulation results of the time-$\tau$ and frequency-$\omega_s$ resolved (a) the intensity change $\Delta I$ and (b) the phase change $\Delta\phi$ in the case of resonance $\omega_0 = 2.35$ PHz, $\Gamma=100$ PHz, and $\gamma=0.000001$ PHz. The horizontal axis represents the time delay ($\tau$) of the probe pulses with respect to the pump pulses. The vertical axis represents the frequency of the probe pulse. The maximum and minimum values of the phase changes are 0.5 and -0.5 rad, respectively.

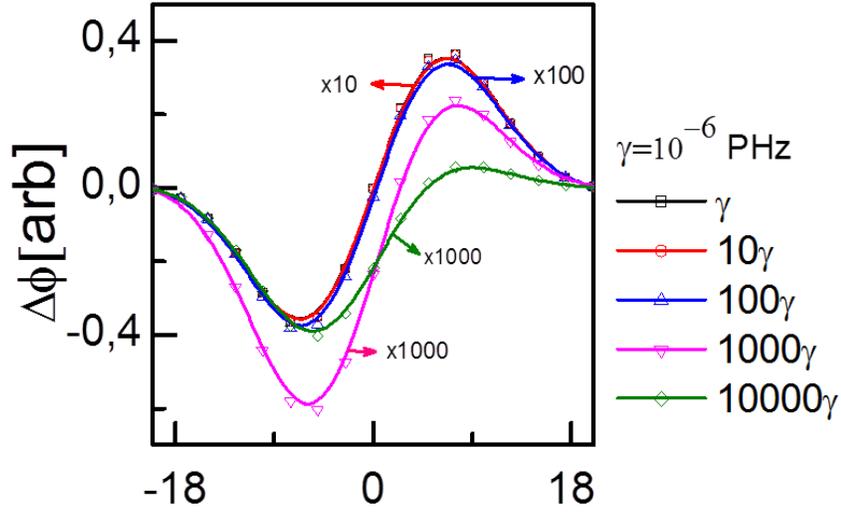

Fig. 3 Population rate dependence of the phase change ($\Delta\phi$). Time delay ($\tau$) dependent phase change at $\omega_s = 2.5$ PHz in various population rates. The open squares, open circles, open triangles, open inverted triangles, and open diamonds correspond to $\gamma_1 = \gamma$, $\gamma_2 = 10\gamma$, $\gamma_3 = 100\gamma$, $\gamma_4 = 1000\gamma$, and $\gamma_5 = 10000\gamma$ respectively. Here, $\gamma = 10^{-6}$ PHz. The same sequenced magnitudes of phase changes were multiplied by 1, 10, 100, 1000, and 1000 numbers.

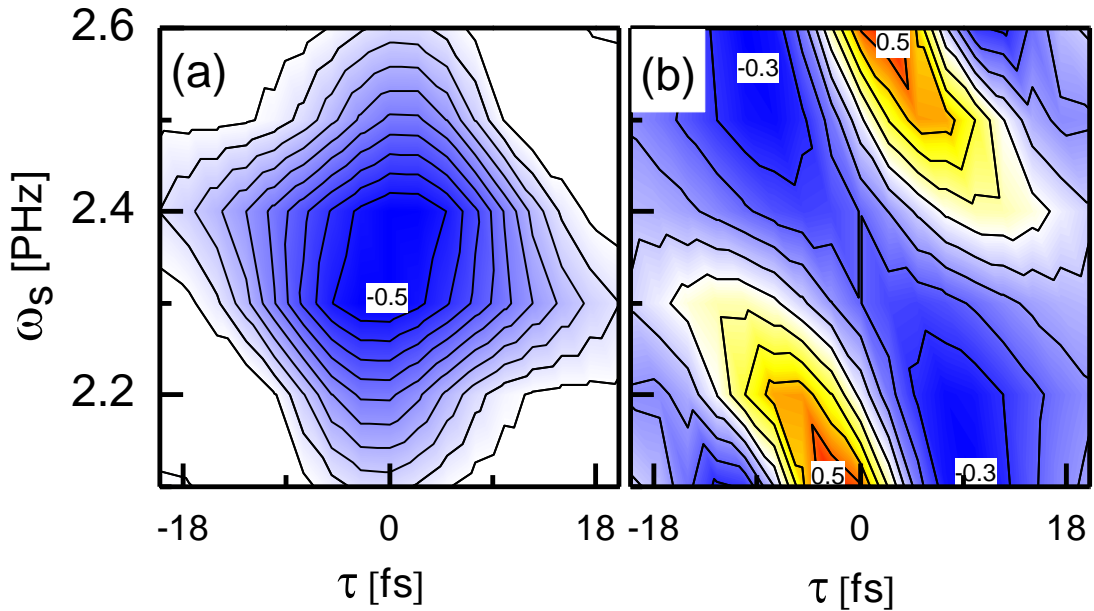

Fig. 4 The frequency-$\omega_s$ and time-$\tau$ resolved (a) the intensity change $\Delta I$ and (b) the phase change $\Delta\phi$ for the chirped probe-pulse. The band gap of the sample is same as the center frequency ($\omega_0 = 2.35$ PHz) of the incident pulse.